# A 3D biomechanical vocal tract model to study speech production control: How to take into account the gravity?

### Stéphanie Buchaillard[1,2*], Pascal Perrier[1*] & Yohan Payan[2*]

[1]ICP – UMR CNRS 5009, INPG & Univ. Stendhal, Grenoble, France.

[2]TIMC – UMR CNRS 5525, Univ. Joseph Fourier, Grenoble, France

buchail@icp.inpg.fr, perrier@icp.inpg.fr, Yohan.Payan@imag.fr

***Abstract.*** *This paper presents a modeling study of the way speech motor control can deal with gravity to achieve steady-state tongue positions. It is based on simulations carried out with the 3D biomechanical tongue model developed at ICP, which is now controlled with the λ model (Equilibrium-¨Point Hypothesis). The influence of short-delay orosensory feedback on posture stability is assessed by testing different muscle force/muscle length relationships (Invariant Characteristics). Muscle activation patterns necessary to maintain the tongue in a schwa position are proposed, and the relations of head position, tongue shape and muscle activations are analyzed.*

## 1. Introduction

The characteristics of speech production (articulatory and acoustic) signals that carry the linguistic information are the results of a combination of influences related to the properties of the human motor control system, the physical mechanisms underlying the generation of articulatory gestures and sounds, the properties of the human speech perception system, and the constraints of the phonological systems. There are a number of findings supporting the hypothesis that these factors of different kinds interact which each other to determine the "common currency" (Goldstein & Fowler, 2003) shared in perception and production of speech. In this framework we have been now working for many years on a model of speech production called *GEPPETO*[1] in order to quantitatively assess the respective influences of target based optimal motor control strategies and of biomechanical properties of the vocal tract articulators.

In the control model defined in *GEPPETO* (Perrier et al., 2005), movement targets are determined by perceptual constraints: they correspond to regions of the formant space in which sounds are perceptually strictly equivalent. Motor control variables are muscle recruitment thresholds, as suggested by the λ model (Equilibrium-Point Hypothesis, Feldman, 1986). These variables specify the equilibrium state of the articulators. For a given phoneme sequence, the choice of the motor control variables associated with each phoneme is determined by an optimal planning that selects the target motor control variables ensuring that the desired target regions of the formant space are reached, while minimizing a speaker-oriented criterion (currently the length of the path in the motor control space). Speech transitions between successive elementary sounds are generated by the shift of the motor control variables at a constant rate between successive target commands. The physical modeling currently proposed in *GEPPETO* includes 2D (Payan & Perrier, 1997) and 3D (Gérard et al., 2006) biomechanical models of the tongue, as well as an acoustic analog of the vocal tract. In this model, the actual movements of the tongue, and then the temporal characteristics of the acoustic signal are the result of the combination of the effects of the time variation of the control variables and of the biomechanical properties of the tongue.

---

[*]This project is supported by the EMERGENCE Program of the Région Rhône-Alpes and by the P2R Program funded by the CNRS and the French Foreign Office (POPAART Project)

[1] GEPPETO holds for "**GE**stures shaped by the **P**hysics and by a **PE**rceptually **O**riented **T**argets **O**ptimization".

With *GEPPETO,* using the 2D biomechanical model of the tongue, it was possible to show that complex kinematic patterns, such as multi-peaks velocity profile (Payan & Perrier, 1997) and loops-like articulatory paths (Perrier et al., 2003) could be largely due to biomechanical influences, including muscle anatomy, soft tissues deformation and tongue palate interaction. In this paper a study of the impact of the gravity on tongue positioning is proposed on the basis of simulations carried out with the 3D biomechanical model of the tongue. Consequences on the adjustment of motor control variables are considered. The importance of the gravity in speech articulators positioning has been demonstrated by different studies (e.g. Shiller et al., 1999; Tiede et al., 2000). It has in particular applications for the analysis of speech production using MRI anatomical images of the vocal tract.

First, the recent improvements of the 3D biomechanical tongue model are presented. In addition to the integration of the gravitational field, these improvements include changes in elastic characteristics of tongue tissues, the account of contacts between tongue and palate and teeth, a modeling of the links between hyoid bone and other bony structures, and the implementation of the λ model. Using this model the potential impact of the gravity on tongue shape in the absence of muscle activation is evaluated, and the importance of a proper account of short-delay sensory feedback is shown, in order to counteract the effect of gravity. Basic muscle activation patterns are proposed that permit to maintain a schwa position when the head is in vertical position.

## 2. The 3D tongue model and related bony structures

### 2.1     Geometrical modeling

The 3D vocal tract model was originally developed by Gerard et al. (2003, 2006). It included biomechanical models of the jaw, the teeth, the palate, the pharynx, the tongue and the hyoid bone. The geometries of these models were extracted from data collected on a human speaker (plaster cast of the palate optically scanned, MRI images and CT scans). The tongue and the hyoid bone were represented by volumes discretized with 3D elements while the other structures were modeled by surfaces discretized with 2D elements. The arrangement of the 3D elements inside the tongue volume was designed in order to identify muscular structures. Eleven muscles were therefore modeled through individual sets of adjacent elements. Insertions on the hyoid bone of tongue muscles and of four external muscles were modeled. External muscles are represented by a set of springs.

## 2.2    Mechanical modelling

2.2.1 Tongue biomechanics

The tongue was modeled as a soft tissue attached to the hyoid bone, the jaw and the styloid process, in interaction through soft contacts with the anterior part of the palate and with the lower teeth. It consists of a 3D Finite Element model assuming a large deformation framework with a non-linear hyper-elastic constitutive law (Gérard et al., 2005, 2006). Arguing the fact that the constitutive law (i.e. the stress-strain relationship) proposed by Gérard et al. (2005) was extracted from data collected on tongue tissues removed from a fresh cadaver, our group proposed to adapt this original law to try to distinguish between passive tissues and tissues that belong to active muscles. Therefore, two different constitutive equations were introduced: one describing the passive behavior of the tongue tissues and another one modeling the stiffness of the tissues as an increasing function of tongue muscles activations. For a given element of the model, the passive (respectively the active) constitutive law is used if the element belongs to a region that is supposed to be passive (respectively active). The passive constitutive law was directly derived from the non-linear law proposed by Gerard et al. (2005). However, as the stiffness of passive tissues removed from a cadaver is known to be lower than the one measured on in vivo tissues, it was decided to modify the constitutive law originally proposed in our tongue model. To our knowledge, one of the most relevant in vivo measurements provided in the literature for human muscle stiffness is the one proposed by Duck (1990). This work provided a value of 6.2kPa for a human muscle in its rest position, and a value of 110kPa for the same muscle when it was contracted. As the original constitutive law of Gérard et al. (2006) provided a value of 1.15kPa for the stiffness at the origin (i.e. the Young modulus value at low strains), it was decided to multiply by a factor of 5 this original constitutive law, in order to reach the 6.2kPa value at the origin while maintaining the overall non-linear shape of the passive constitutive law. Figure 1 plots the corresponding law.

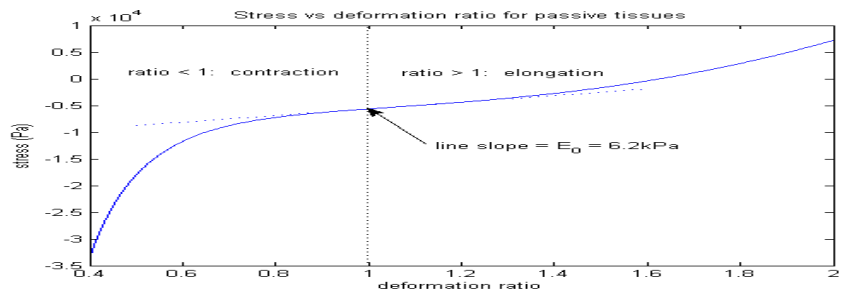

**Figure 1.** Stress/Strain constitutive law for passive tissues (Yeoh second order material with $C_{10}$=1037Pa and $C_{20}$=486Pa). The curve's tangent for a deformation rate equals to 1 gives the Young modulus at low strains.

As concerns the constitutive law that describes the mechanical behavior of an element belonging to an active muscular region, it was decided to arbitrarily multiply the passive constitutive law by a factor that is a function of the muscle activation. The underlying idea is that an activation of the muscle leads to an increase of its stiffness. The multiplying factor was chosen in order to maintain the stiffness value below 110kPa when maximal muscle activation is simulated.

The tongue mass density was chosen equal to 1000kg/m$^3$. With the proposed finite element mesh the tongue mass was calculated equal to 92g. The damping was defined by specifying mass and stiffness Rayleigh damping constants that were set respectively to 10 and 0.7.

2.2.2 Hyoid bone biomechanics

The hyoid bone is represented by 4-nodes 3D tetrahedral elements. Its mass approximates 5g. Its body gives insertion to the mylohyoid, the geniohyoid and the posterior part of the genioglossus, and the greater cornua to the hyoglossus. Nodes on the tongue mesh corresponding to muscle insertions were selected as insertion nodes. A set of 10 springs emerging from the hyoid bone were used to represent the anterior part of the digastric, its posterior part, as well as the sternohyoid, omohyoid and thyrohyoid muscles. The spring's stiffness was set to 25N.m$^{-1}$. The insertions nodes and the hyoid bone define a rigid region to ensure a realistic coupling between the hyoid bone and the surrounding muscles.

2.2.3 Contacts between tongue and hard structures

Three contact regions were determined to represent the contact between the tongue and 1) the hard palate, 2) the soft palate, 3) the lower teeth and the inner face of the mandible. For each region, a group of elements belonging to the tongue surface, called the contact surface, and to the second region of interest, called the target surface, were selected. At each step of the resolution, collisions and contacts were detected for every surface/target pair and resolved using an augmented Lagrangian contact algorithm (iterative penalty method). We used a contact stiffness factor of 0.10 (Ansys FKN parameter) for the different pairs, and an allowable penetration (FTOLN parameters) of 0.2mm for the first two pairs (contact with the palate), and 0.3mm for the last one.

## 3. Muscle force generation mechanisms based on the λ model

### 3.1 Implementation of the λ model

Each tongue muscle was represented by a set of adjacent elements inside the mesh. A *main fiber direction* was defined for each muscle within the corresponding set of elements. For each muscle; its activation was functionally modeled by an external generator (based on the Equilibrium Point Hypothesis, see below) applying distributed forces along *macro-fibers* joining the nodes of the corresponding finite elements set along the main fiber direction. Several macro-fibers were defined for each muscle. For more details about the distribution of the force along the macro-fibers, see Payan & Perrier (1997).

Force generation mechanisms are modeled with the λ model (Feldman, 1986), in which muscle activation $A$ depends both on central commands (λ's) and on feedback information concerning the muscle length $l$ and its variation rate $\dot{l}$, according to the equation:

$$A = [l - \lambda + \mu\dot{l}]^{+} \text{ with } \mu = 0.01\text{s} \qquad (1)$$

The active muscle force $\tilde{M}$ associated with this activation is modeled as

$$\tilde{M} = \rho[\exp(cA) - 1] \qquad (2)$$

where ρ is a parameter accounting for the force generation capability of the muscle. This capability is linked to the cross-sectional area of the muscle and it was fixed to the values given in Table 1, on the basis of Payan & Perrier (1997) and Van Eijden et al.

(1997). The c parameter was adjusted specifically for the 3D model, as will be explain below in section 3.2.

A unique central λ command is defined for each muscle. However, given the model of force distribution over macro-fibers, it is necessary to infer from the main central command λ, a motor command $λ_i$ adapted specifically for each macro-fiber. To do so, the lengths of the macro-fibers were calculated in the rest position, and the $λ_i$ values were calculated with each muscle on the basis of these lengths. For the longest fiber, $λ_i$ was fixed to the central command λ, and for the other macro-fibers $λ_i$ was calculated as ($w_{i*}λ$), where $w_i$ is the ratio of the length of the macro-fiber divided by the length of the longest macro-fiber of he muscle ($w_i$ <1). This approach was chosen, in order to distribute the force evenly over the macro-fibers of a same muscle, when tongue is at rest. The values $w_i$ are constant values independently of tongue position.

| GGp | GGm | GGa | Stylo.. | GenioH.. | Mylohyoid | Hyo | Vert | Trans | IL | SL |
|-----|-----|-----|---------|----------|-----------|-----|------|-------|----|----|
| 29  | 26  | 12  | 24      | 18       | 39        | 65  | 30   | 45    | 19 | 14 |

**Table 1.** ρ factor specifying muscle force generation capabilities of each muscle described in the tongue model.

Following Laboissière et al. (1996) approach, and in agreement with Payan & Perrier (1997), a model of the dependence of muscle force on muscle variation rate was also taken into account. According to this model, muscle force decreases dramatically, when length variation rate becomes too large.

3.2    Choice of feedback gain and delay

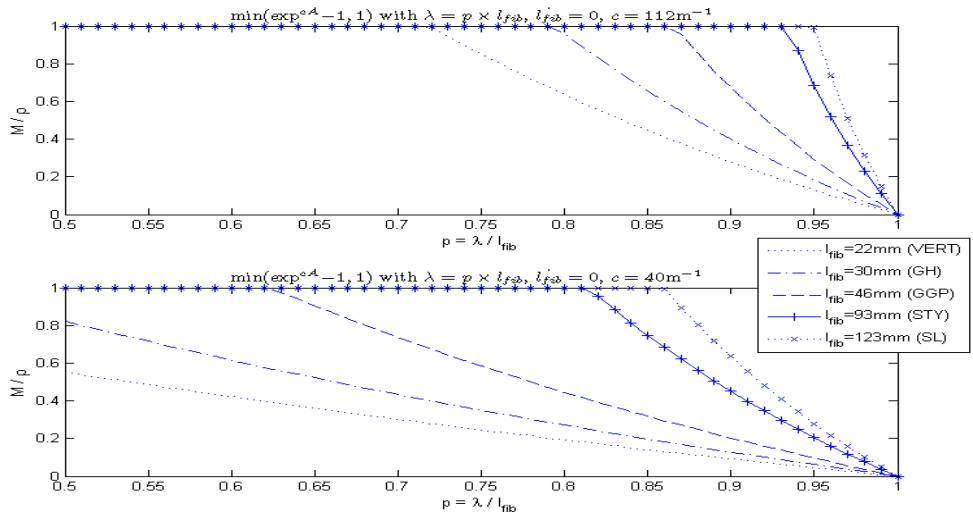

**Figure 2:** Variation of normalized muscle activation with respect to the ratio of the central command λ and the fiber length $l_{fib}$, for different muscles. Top panel: with c=112m$^{-1}$; bottom panel : c=40m$^{-1}$

Our first attempt to implement the λ model in our 3D tongue model is based on a value of 112m$^{-1}$ for the c parameter in Equation 2. This value was proposed by Laboissière et al. (1996) for their 2D biomechanical jaw model. It was also used in the 2D tongue models of Payan & Perrier (1997) and Perrier et al. (2003). Interestingly while these 2D biomechanical models of the tongue were all stable, strong instabilities were observed in the 3D model with and without central muscle activation. The top panel of Figure 2 provides an explanation for this instability. This figure plots the normalized muscle activation with respect to the ratio of the central command λ and the fiber length $l_{fib}$, for different muscles. It can be observed that a small variation in the muscle activation can lead to dramatic changes in the muscular activation level. Therefore, the value of the c parameter was decreased down to 40 $m^{-1}$, which corresponds to a fair compromise between the level of reflex activation and stability (bottom panel).

Different tests were also made with the duration of the feedback delay Δ necessary to take into account the muscle length and its variation rate. A delay in the range 10 to 20ms does not generate any instability.

## 4     Influence of orosensory feedback on posture stability

### 4.1     Influence of the Gravity in absence of feedback

The tongue models originally provided by our group (Payan & Perrier, 1997; Perrier et al., 2003; Gerard et al., 2003, 2006) did not include the gravity since this was not straightforward at all. Indeed, one must be aware that the tongue geometry extracted from medical images of the speaker (CT scans and MRI images) measures the rest position of the tongue *with* the gravity. Since this geometry is usually taken as an input for the initial shape of the tongue mesh, this mesh should be "pre-stressed", i.e. should already include the internal stresses that are only due to the gravity. To our knowledge, most of the tongue models published in the literature do not model this pre-stress. Therefore, this study aims at investigating more in details the way the control model implemented in GEPPETO accounts for the gravity, in normal conditions (i.e. with the head in the vertical plane) and in disturbed ones (the head in the horizontal plane).

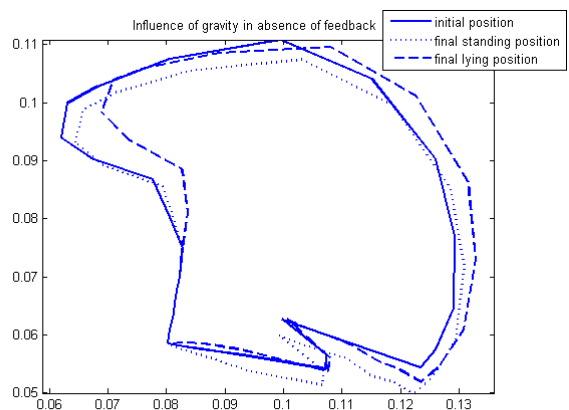

**Figure 3.** Influence of gravity without the control model implemented in GEPPETO. The contours are represented in the midsagittal plane after 1s of simulation. The solid, dotted and dashed contours correspond respectively to the initial tongue position, the final position for a standing subject and the final position a lying subject. Units of the X- and Y-axis are m

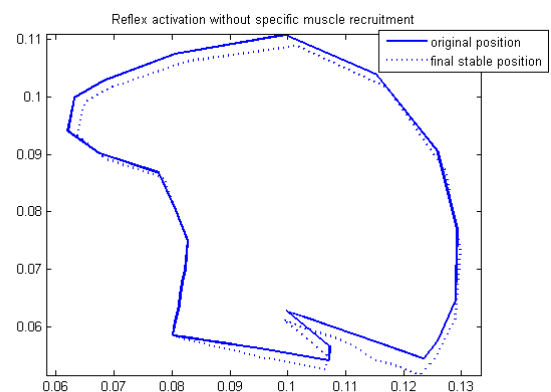

**Figure 4.** Influence of gravity with reflex activation (λ model) and without any centrally specified muscle activation. The solid contour corresponds to the initial tongue position and the dotted contour to the final position for a 1s simulation. Units of the X- and Y-axis are m

As a first step, tongue movement was simulated during one second, in the absence of any feedback and of any centrally controlled muscle activation. In this aim, the λ model was deactivated and two conditions were tested: one with head in vertical position comparable to the head of a person sitting or standing, and one with the head in the horizontal plane similar to the head position when subjects are lying.

Figure 3 plots the tongue midsagittal contours for both simulations: a standing subject (dotted contour) and a lying subject (dashed contour). In the standing condition, tongue is deformed in the backward and downward direction towards the larynx, leading to an rest lingual shape, which tends to obstruct the lower part of the pharynx. In the lying condition, the displacement is even larger towards the back of the mouth, with a tongue that is also going to obstruct the pharyngeal tract. It is here interesting to note that this fall down towards the larynx is often observed by surgeons with anaesthetized patients (i.e. deprived from any controlled muscular activity) lying on the operating table.

## 5      Muscle activations patterns in schwa position

In order to stabilize the initial tongue position, defined as the schwa position in our model, the reflex activation alone in absence of specific muscle recruitment was found to be insufficient (see Figure 4). As shown in Figure 5, the combined activation of the mylohyoid (MH), which stiffens the floor of the mouth and consequently elevates the tongue body, and of the posterior genioglossus (GGP), which pulls the tongue forward, seems to be satisfactory to maintain the initial position for a standing subject (dotted contour). The λ command for GGP was set to 97% of the GGP rest length, and for the MH it was set to 90% of the GGP rest length. However, for a lying subject, the same activation pattern is not able to fully counteract the influence of gravity. The maximum force levels reached during the simulation (sum of the forces applied on the different fibers) equaled 1.06N for the GGP and 2.99N for the MH for a standing position, and 1.33N for the GGP and 2.66N for the MH for a lying position.

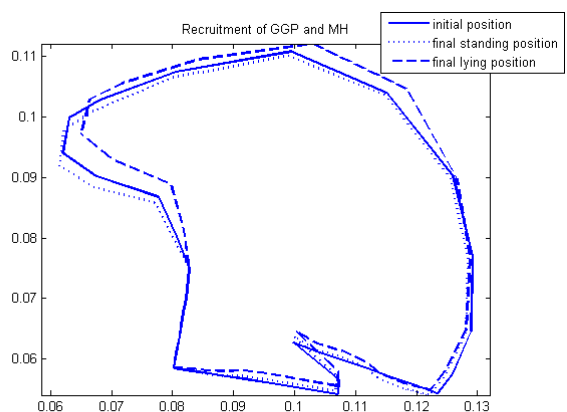

**Figure 5.** Influence of gravity with reflex activation (λ model) and without the centrally specified activation of GGP and MH (see text). The solid contour corresponds to the initial tongue position and the dotted contour to the final position for a 1s simulation. The contours are represented in the midsagittal plane after 500ms of simulation. The solid, dotted and dashed contours correspond respectively to the initial tongue position, the final position for a standing subject and the final position a lying subject. Units of the X- and Y-axis are m.

## 6      Conclusions

The potential influence of gravity on articulatory movements and on their control during speech production has not been much studied yet, in spite of the fact that MRI data provide evidence that the tongue shape can be influenced by changes in the speaker's head orientation. The modeling framework proposed in *GEPPETO,* including a realistic 3D biomechanical model of the tongue and a control model based on the Equilibrium-Point Hypothesis, has permitted to quantitatively assess (1) how the integration of short-delay orosensory feedback helps to counteract the effect of gravity, (2) to which extent muscle recruitments learned to control a tongue posture in a given head position are adapted to the control of the posture in another head orientation. It gives the possibility to study in future works to which extent the speech motor control system has to integrate the gravity to achieve speech movements with enough accuracy.